\documentclass[aps,twocolumn]{revtex4}
\usepackage{graphicx}

\begin{document}

\title{Sub-50 mK electronic cooling with large-area superconducting tunnel junctions}

\author{H. Q. Nguyen}
\affiliation{Low Temperature Laboratory (OVLL), Aalto University School of Science, P.O. Box 13500, 00076 Aalto, Finland}
\affiliation{Nano and Energy Center, Hanoi University of Science, VNU, Hanoi, Vietnam}
\author{M. Meschke}
\affiliation{Low Temperature Laboratory (OVLL), Aalto University School of Science, P.O. Box 13500, 00076 Aalto, Finland}
\author{H. Courtois}
\affiliation{Universit\'e Grenoble Alpes, Institut N\'eel, F-38042 Grenoble, France}
\affiliation{CNRS, Institut N\'eel, F-38042 Grenoble, France}
\author{J. P. Pekola}
\affiliation{Low Temperature Laboratory (OVLL), Aalto University School of Science, P.O. Box 13500, 00076 Aalto, Finland}

\begin{abstract}
In electronic cooling with superconducting tunnel junctions, the cooling power is counterbalanced by the interaction with phonons and by the heat flow from the overheated leads. We study aluminum-based coolers that are equipped with a suspended normal metal and an efficient quasi-particle drain. At intermediate temperatures, the phonon bath of the suspended normal metal is cooled. By adjusting the junction transparency, we control the injection current, and thus temperature of the superconducting lead at the optimum cooling point. The best device shows remarkable cooling from 150 mK down to about 30 mK, a factor of five in temperature at a power of 40 pW. We discuss heat transport in our device and the reasons for cooling saturation at the low temperature end. 
\end{abstract}
\maketitle

\section{Introduction}
When a Normal metal - Insulator - Superconductor (NIS) junction is biased near the superconducting gap, hot electrons tunnel into the superconductor and the normal metal is cooled down \citep{NahumAPL,MuhonenRPP, GiazottoRMP06, PekolaPRL04, LeivoAPL96,VasenkoPRB10}. Based on this effect, micron-size electronic coolers attached to a dielectric platform \citep{UllomAPL13} can cool an external object at temperatures well below 300 mK, a regime that traditionally belongs to macroscopic cryostats, such as adiabatic demagnetization or 3He-4He dilution refrigerators. In the best demonstration to date \citep{MillerAPL08}, a superconducting transition edge sensor has been cooled on a platform attached to NIS junctions from 300 to 200 mK. Clearly, such a platform is of great interest as an alternative method to bring ultra-sensitive low temperature detectors \citep{Enss}, including those at the frontier of astronomical studies of dark matter \citep{DarkMatter}, neutrinos \citep{Neutrino}, or cosmic microwave background \citep{Bicep2}, into their proper working temperature. It allows those devices to work from a bath temperature higher than their own critical temperatures. Moreover, for space borne applications \citep{Tauber}, NIS coolers could reduce the payload noticeably. The ability to reach sub-50 mK regime with NIS coolers is crucial in order to open up new possibilities to cool qubits \citep{MartinisNature10} and nanomechanical resonators \citep{SillanpaaNature13} to their quantum ground states, or to evacuate heat in electron pumping devices \citep{PekolaRMP13}. This fact extends to the next generation of ultra low temperature detectors, like SQUIPT sensors \citep{MMNPhys10} that are even more sensitive once operated in the sub-50 mK temperature regime.

To cool well, the normal metal needs to be isolated from the environment, and at the same time, the hot superconducting leads need to be thermalized with the surrounding bath. In the normal metal, electrons interact with the lattice phonons, with a coupling strength decaying quickly towards low temperatures as $T^5$. Under some conditions, the lattice phonons can decouple from the thermal bath of the substrate phonons \citep{RajauriaPRL07,KoppinenPRL09,PascalPRB13}. In the superconducting leads, hot quasi-particles at an energy just above the energy gap are generated \citep{Schoelkopf14,GiazottoRMP06, PekolaPRL04}. It is a challenge to thermalize them, as quasi-particle relaxation rates decrease exponentially with lowering the temperature. Typically, these hot particles can be evacuated to a quasi-particle trap, which is a layer of normal metal in close contact with the superconductor \citep{ONeilPRB12,Luukanen,CourtPRB08}.

When assuming that the electronic populations in both the normal metal and the superconductor can be described by Fermi distributions at respective temperatures $T_N$ and $T_S$, the cooling power of a NIS junction at its optimum cooling bias $eV\simeq\Delta-0.66k_BT_N$ is given by \citep{GiazottoRMP06}
\begin{equation}
	\dot{Q}_{NIS}\simeq\frac{\Delta^2}{e^2R_T}\left[0.59\left(\frac{k_BT_N}{\Delta}\right)^{3/2}-\sqrt{\frac{2\pi k_BT_S}{\Delta}} e^{-\frac{\Delta}{k_BT_S}}\right].
\end{equation}
Here, $R_T$ is the tunnel resistance, $\Delta$ is the superconducting gap, $k_B$ is Boltzmann constant and $e$ is the electron charge. If the superconducting lead is not properly thermalized so that $T_S (> T_{bath})$ approaches the superconducting transition temperature $T_c$, the term $\exp{-\frac{\Delta}{k_BT_S}}$ becomes significant and $\dot{Q}_{NIS}$ diminishes. In many cases, one assumes no overheating and at low temperature one can neglect the second term, thus $\dot{Q}_{NIS}\propto T_N^{3/2}$. In the low temperature regime $T_N\ll T_c$, the efficiency of the cooler is then given by \citep{GiazottoRMP06}:
\begin{equation}
	\eta=\frac{\dot{Q}_{NIS}}{IV}\simeq0.7\frac{T_N}{T_c}.
\end{equation}
It amounts to about $20\%$ near $T_N$ = 350 mK for aluminum, which is the standard choice of a superconductor. In general, the most significant opposing heat current to $\dot{Q}_{NIS}$ comes from the electron-phonon interaction in the normal metal. The most accepted form for a metal writes
\begin{equation}
	\dot{Q}_{eph}=\Sigma\mathcal V(T_e^5-T_{ph}^5),
\end{equation}
where $\Sigma=2\times10^9$ WK$^{-5}$m$^{-3}$ for Cu, $\mathcal{V}$ is the volume of the normal island, $T_e$ and $T_{ph}$ is the electron and phonon temperature, respectively.

Recently, we have developed a technique \citep{NguyenAPL} to fabricate large-area SINIS coolers targeted at optimizing both $\dot{Q}_{eph}$ and $\dot{Q}_{NIS}$. First, the cooled normal metal is suspended on top of the superconducting electrodes, and thus quite decoupled from the substrate phonons. Second, hot quasi-particles in the leads are efficiently thermalized with a normal metal drain coupled to the superconductor through a transparent tunnel barrier \citep{NguyenNJP}. 

In this paper, we show that these two advanced features, combined with an optimized tunnel junction transparency, improve the performance of a SINIS cooler significantly. At intermediate temperatures where electron-phonon coupling is substantial, phonons in the suspended normal metal are cooled. At low temperature, where the cooler is almost free from electron-phonon interaction, we tune the overheating in the superconducting leads by adjusting the transparency of the cooling junctions tunnel barrier. The most efficient SINIS cooler reaches about 30 mK, which is about $3\%$ of $\Delta/k_B$. We discuss the heat transport and the benefit of having a quasi-particle drain coupled to the superconductor, as well as the possible reasons for the saturation at the lowest temperature.

\begin{table}[b]
\begin{tabular}{@{}ccccccc}\hline
Samples & $P_{O_2}/t_{O_2}$ & $l\times w$ & $d_{Cu}$ & $2R_T$ & Figure\\
& mbar/minutes & $\mu$m $\times$ $\mu$m & nm & $\Omega$ &\\
\hline
A1 & 1.3/5 & 70$\times$4 & 100 & 1.7 & 1, 2, 3 \\
A2 & 1.3/5 & 70$\times$10 & 100 & 0.75 & 1 \\
A3 & 1.3/5 & 70$\times$2 & 100 & 3.4 & 1  \\
A4 & 1.3/5 & 70$\times$4 & 20 & 1.2 & 1  \\
A5 & 1.3/5 & 950 & 100 & 0.8 & 1 \\
B & 9/5 & 70$\times$4 & 60 & 3.7 & 2\\
C1 & 13/5 & 70$\times$4 & 60 & 4.8 & 2, 3 \\
C2 & 13/5 & 70$\times$4 & 60 & 4.6 & 2  \\
D & 50/120 & 70$\times$4 & 60 & 10.5 & 2, 3 \\\hline
\end{tabular}
\caption{\label{table} Parameters of the measured SINIS coolers. $P_{O_2}/t_{O_2}$ refers to the oxidation pressure and time used for producing the tunnel barrier of the cooler. $l \times w$ is length $\times$ width of a rectangular NIS junction. We write the area for A5 in $\mu$m$^2$, as it has an interdigitated shape. $d_{Cu}$ is the thickness of the normal metal island. 2$R_T$ is the tunnel resistance of the two NIS cooling junctions in series. All samples gap $2\Delta$ = 375 $\mu$eV. "Figure" indicates the figure where data on this sample is shown.}
\end{table}

\section{Fabrication and measurement methods}
Following Ref. \citep{NguyenAPL}, the devices are fabricated using photolithography and metal wet etch, see Fig. 1a for a schematic view and inset of Fig. 1b for a top view of the samples. In this work, all coolers have a suspended normal metal bridging two 200 nm thick Al superconducting electrodes sitting on top of a 200 nm thick AlMn quasi-particle drain \citep{NguyenNJP}. The Al and AlMn layers are separated by a thin AlOx layer, oxidized in a mixture of Ar:O$_2$, ratio 10:1 at pressure $2\times10^{-2}$ mbar for 2 minutes, see Table 1 for more sample parameters. All samples except C2 (see below) have in addition (to the drain) a quasi-particle trap of Cu next to the junction. Coolers are measured with standard four-probe technique in a dilution cryostat. The electronic temperature $T_N$ on the normal metal is measured by using a pair of smaller NIS junctions \citep{GiazottoRMP06}, where the voltage drop under a constant current (about 1 nA) is calibrated to the cryostat temperature. Below the lowest cryostat temperature of 50 mK, the temperature $T_N$ is extracted based on an extrapolation of BCS theory using data from the higher temperature regime. The measurement noise-related uncertainty is less than 300 $\mu$K, see Fig. 2. Extracting the temperature from fitting the current-voltage characteristic to isothermal theory curves \citep{RajauriaPRL07,ONeilPRB12} gives identical results, including those below the lowest temperature of the cryostat where the difference is smaller than 1 mK between the two methods.

\section{Experimental Results}

\begin{figure}[bt]
\begin{center}
\includegraphics[width=\columnwidth,keepaspectratio]{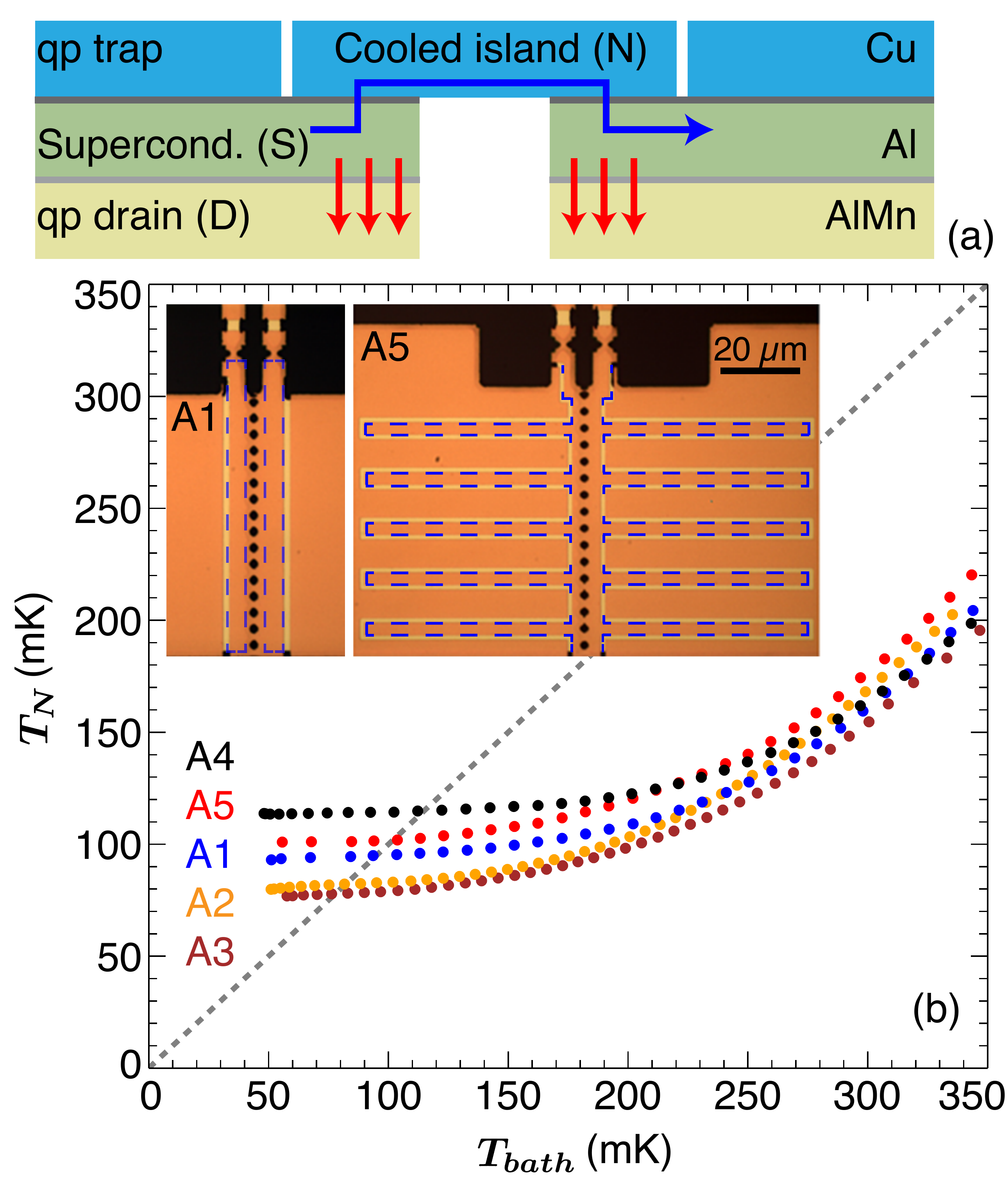}
\caption{(a) Schematic cross-sectional view of SINIS cooler, where Cu is for the normal metal island and quasi-particle traps, Al for the superconducting leads, and AlMn for the quasi-particle drains. The blue arrow indicates the charge current and the red arrows show the heat current. (b) Normal metal electronic temperature $T_N$ reached at the optimum bias vs bath temperature $T_{bath}$ for samples made on the same wafer but differing in their geometries. Compared to the standard sample A1 (70$\times$4 $\mu$m$^2$ for one junction), A2 has twice the junction area, A3 has half the junction area, and A4 has 20$\%$ of its normal metal volume. The left inset shows an image of the standard sample A1, and the right inset shows an image of A5 with an interdigitated junction shape. The NIS junctions are bordered with dashed lines.}
\label{elph}
\end{center}
\end{figure}

Figure 1b demonstrates that, compared to the cooling power, electron-phonon interactions are negligible in our devices. Here, all coolers are made of the same AlMn/AlOx/Al/AlOx/Cu multilayer from a single wafer, but with a varying junction geometry. Taking A1 as the reference sample, A2 junction has a double junction size \citep{footnoteA2}, and A3 half of that in A1. A4 has the same junction size as A1, but its Cu thickness is 20$\%$ of the value in A1. The junction in sample A5 has an interdigitated shape, where quasi-particle traps surround the NIS junctions. As $\dot{Q}_{NIS}\propto R_T^{-1}$ and $\dot{Q}_{eph}\propto\mathcal{V}$, varying the junction size and the normal metal volume can give information about the heat balance. Nonetheless, it is hard to see a real trend in the data, even near 300 mK where $\dot{Q}_{eph}\sim \Sigma\mathcal{V}T_{ph}^5$ is expected to be significant \citep{footnoteA4}. We conclude the present electronic coolers are almost free of electron-phonon interaction so that $\dot{Q}_{eph}$ is not responsible for the saturation of the normal metal temperature $T_{N,min}$ in the low temperature end. As a consequence, adjusting the N metal geometry does not improve its low-temperature performance.

\begin{figure}[t]
\begin{center}
\includegraphics[width=\columnwidth,keepaspectratio]{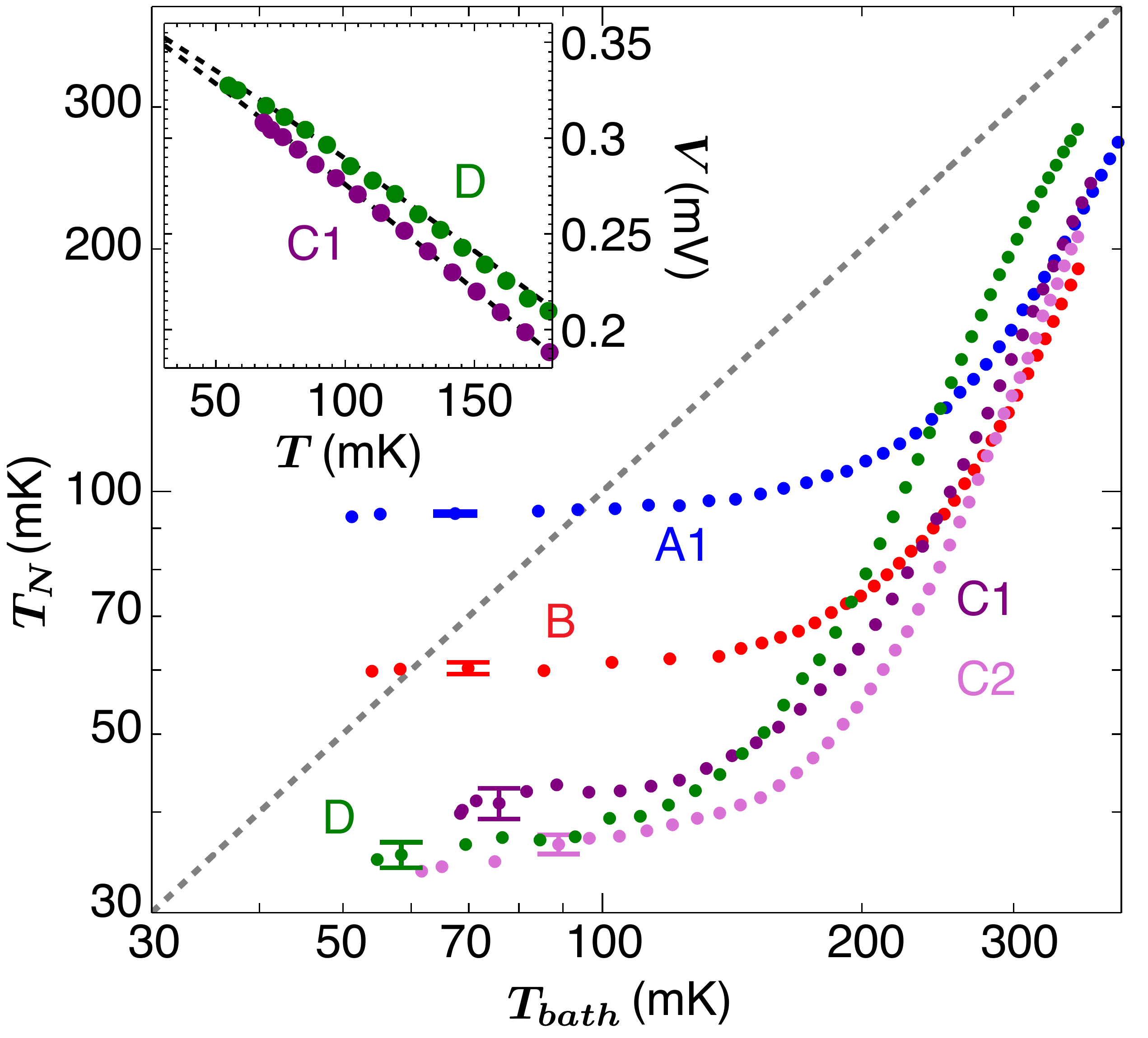}
\caption{Temperature of the normal metal island $T_N$ at the optimum bias as a function of bath temperature $T_{bath}$. Samples A1, B, C1, and D differ only in their tunnel resistances $R_T$ (Table 1). C2 is an improved version of C1, see text. The gray dotted line is the 1-1 line at the boundary between cooling and heating. The error bar for each data set from the measurement is represented. The inset shows thermometer calibrations, \mbox{i.e.} voltages of the probing junctions at a current of 1.5 nA for samples C1 and D at cryostat temperatures below 200 mK. Dots are experiment data and dashed lines fit to BCS theory.}
\label{lowT}
\end{center}
\end{figure}

Our main practical achievement is presented in Fig. 2: starting from a 150 mK bath temperature, the most powerful coolers reach a 30 mK electronic temperature, a five-fold reduction of temperature. Samples A1, B, C1, and D are made using an identical recipe and differ only in the cooler barrier resistance $2R_T$ with respective values 1.7, 3.7, 4.8, and 10.5 $\Omega$. A smaller $R_T$ leads to a larger cooling power, which is beneficial. Nevertheless, it also leads to a stronger quasi-particle injection, which overheats the superconducting leads and degrades cooling at low temperature. Adjusting the tunnel resistance $R_T$ is therefore essential for optimizing electronic cooling. Within our sample set, sample A1 with a large cooling power works best near 300 mK, but saturates at 94 mK. Sample B reaches a lower temperature of 60 mK thanks to a smaller dissipation by injection current. Sample D has the lowest cooling power at high temperature, but cools down to 32 mK. Sample C1 is a compromise between B and D and performs well over a wide temperature range. We conclude that the higher cooling power that is desirable at high temperatures compromises the performance of the device at the low-temperature end, due to the back-flow of heat from the overheated leads.

In order to investigate the limits of electronic cooling in the samples, we have improved sample C1 in a number of ways, so that it is afterwards called C2. First, C2 was equipped with a pair of direct quasi-particle traps, which locate 1 $\mu$m away from the junction. The superconducting leads are then affected by the direct contact with the normal metal, and are thus expected to conduct heat better \citep{CourtPRB08,Luukanen}. This would improve the performance of C2 if the quasi-particle drain was limiting the performance in C1. Second, we bonded sample C2 with Au wires, which have a much higher thermal conductance as compared to the usual superconducting Al wires employed in other samples. Finally, we measured sample C2 in a rf-tight double-shielded sample stage \citep{SairaPRB12}. These improvements work toward eliminating extra heating from quasi-particles, phonons, as well as the radiation from the high temperature parts of the cryostat. Despite of all the effort, sample C2 performs only slightly better than C1 in the whole temperature range covered. This implies that the drain constitutes by itself an efficient quasi-particle trap that crucially helps an electronic cooler to reach 30 mK.

\begin{figure}[bt]
\begin{center}
\includegraphics[width=\columnwidth,keepaspectratio]{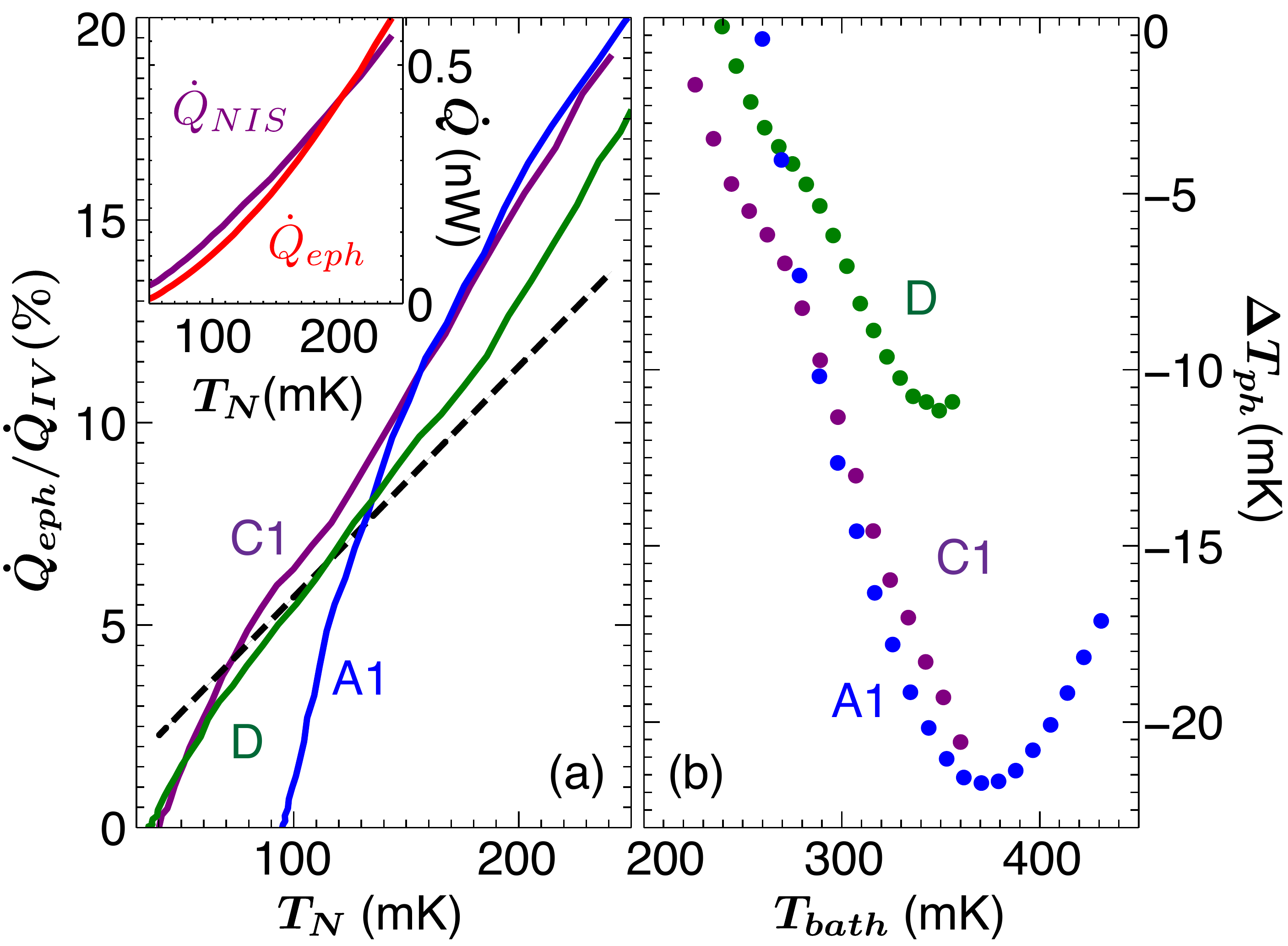}
\caption{(a) Apparent efficiency calculated with the assumption of metal phonons thermalized at the bath temperature for samples A1, C1, and D compared to the theory prediction Eq. (2) (black dashed line). The inset shows the calculated $\dot{Q}_{NIS}$ when assuming $T_S=T_{bath}$ and $\dot{Q}_{eph}$ when assuming $T_{ph}=T_{bath}$ for sample C1. (b) Extracted phonon temperature of the normal island $\Delta T_{ph}=T_{ph}-T_{bath}$ assuming the theoretical efficiency and no overheating of the leads.}
\label{power}
\end{center}
\end{figure}

Let us now compare the cooler performance to theoretical predictions. We first assume a good thermalization of the metal phonons to the bath temperature: $T_{ph}=T_{bath}$. Figure 3a shows the efficiency $\dot{Q}_{eph}/IV$ calculated within this framework for samples A1, C1 and D at the optimum cooling point. The measured quantity exceeds the prediction of Eq. (2) over a wide temperature range. In the inset, the comparison of the NIS cooling power Eq. (1) with the electron-phonon coupling power Eq. (3) confirms this conclusion for sample C1: the two curves cross near $T_N$ = 220 mK, above which $\dot{Q}_{eph}>\dot{Q}_{NIS}$. Here we assumed no overheating in the superconductor: $T_S=T_{bath}$ in Eq. (1), which gives an upper estimate for $\dot{Q}_{NIS}$. As the excess efficiency occurs at intermediate temperatures near $T_{bath}$ = 300 mK, it is best explained by assuming that not only the electrons but also the phonons of the normal island cool, \mbox{i.e.} $T_{ph}<T_{bath}$. Our samples are particularly suited to this to take place as the normal metal island is suspended on the superconducting electrodes and thus decoupled from the substrate. In order to estimate the phonon cooling, we calculated the drop in phonon temperature $\Delta T_{ph}=T_{bath}-T_{ph}$ necessary to fulfill the heat balance Eq. (2), see Fig. 3b. A phonon cooling of about 20 mK is obtained in samples A1 and C1. Again, the probable over-estimation of $\dot{Q}_{NIS}$ makes that the present phonon temperature estimation is a minimum value.

\section{Analysis of the results and discussion}
\begin{figure}[bt]
\begin{center}
\includegraphics[width=\columnwidth,keepaspectratio]{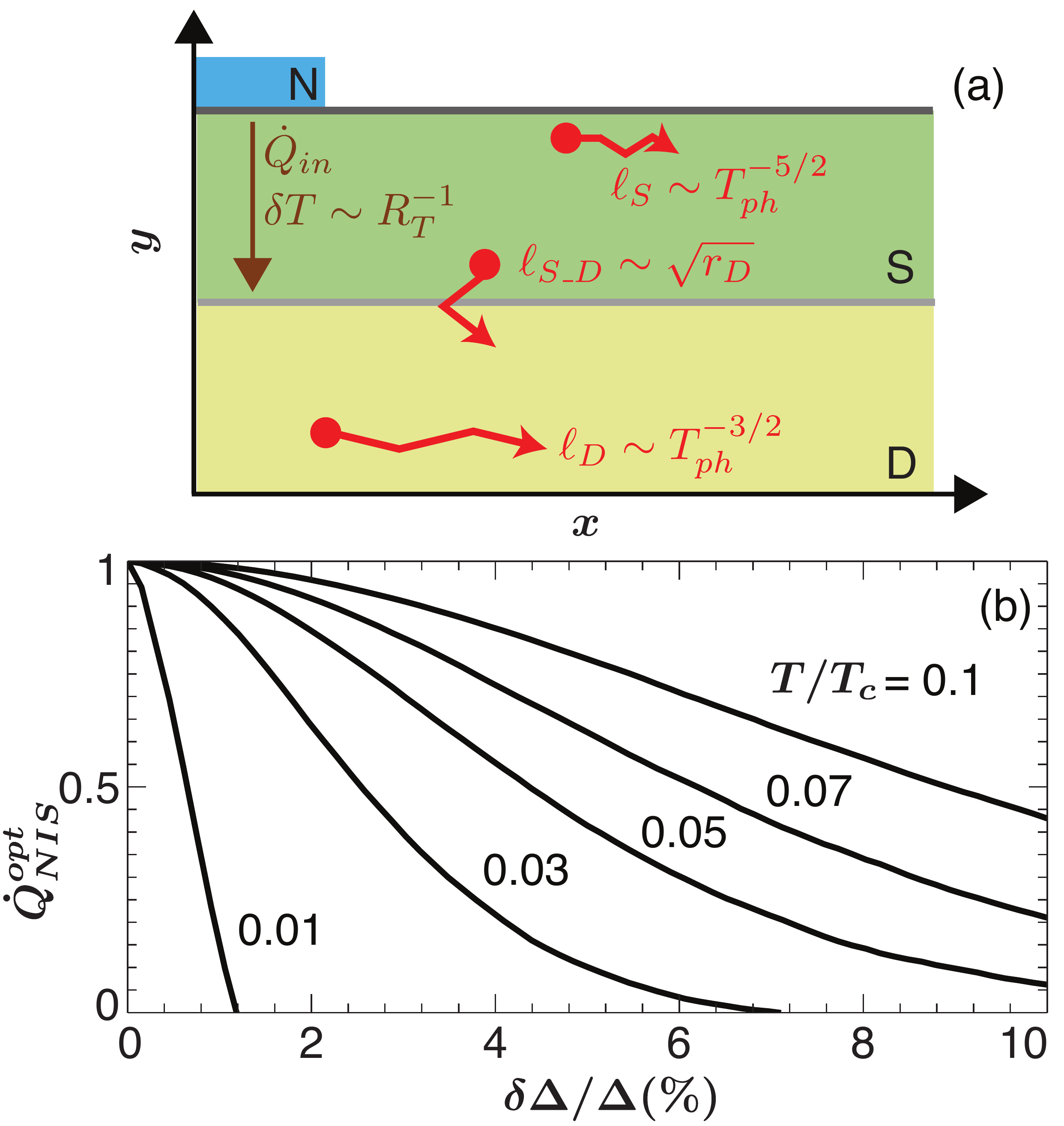}
\caption{(a) Sketch of the thermal transport when the superconductor couples to the quasi-particle drain. The red arrows represent the heat current where quasi-particles relax. We label their most notable dependences, see text. (b) Optimum cooling power normalized by that of an ideal uniform junction as a function of the gap non-uniformity at different temperatures.}
\label{dgap}
\end{center}
\end{figure}

To illustrate the benefit of having an efficient quasi-particle drain, we now analyze the heat transport using a diffusion model sketched in Fig. 4a. In the $y$ direction, the temperature change $\delta T_S(y)$ across the superconductor thickness $d_S$ is estimated to be $\frac{y\dot{Q}_{in}}{A\kappa_S}$, where $\kappa_S$ is the electronic thermal conductivity of the superconductor and $\dot{Q}_{in}$ is the input power from the cooler barrier of area A. The temperature drop is only $\delta T_S=0.4$ mK under a 1 nW input power at 250 mK, and thus negligible. Along the $x$ direction, we use a thermal conductivity $\kappa_D=\mathcal{L}_0\rho_D^{-1} T_D$ in the normal metal drain, where $\mathcal{L}_0$ is the Lorenz number, $T_D$ the local temperature and $\rho_D$ the resistivity. The density of the heat current is $\dot{q}_D=\Sigma_D (T_D^5-T_{ph}^5)$, where $\Sigma_D$ is the electron-phonon coupling in the drain material. For a small temperature change, we obtain a heat relaxation length $\ell_D = \sqrt{\frac{\mathcal{L}_0}{5 \rho_D\Sigma_D}} T^{-3/2}_{ph}$. For AlMn, this yields $\ell_D$=11 $T^{-3/2} \mu$mK$^{-3/2}$ (120 $\mu$m at 200 mK) \citep{footnote1}. In the superconducting leads, the heat conductivity is decreased by a factor $6(\frac{\Delta}{k_BT_S})^2 \exp(-\Delta/k_BT_S)$ while the heat exchange is reduced by $\exp(-\Delta/k_BT_S)$ \citep{Timofeev}. We obtain $\ell_S=\frac{\Delta}{\pi k_B}\sqrt{\frac{6\mathcal{L}_0}{5 \rho_{S}\Sigma_S}} T^{-5/2}_{ph}$. For pure Al without a trap, $\ell_S$=50 $T^{-5/2} \mu$mK$^{-5/2}$ (3 mm at 200 mK). The heat transport along the superconducting leads is thus much less efficient than in the drain, as expected.

When the superconductor couples to the drain, the heat flux through the barrier in between is given by $\dot{q}=\frac{\sqrt{2\pi k_B\Delta^3}}{e^2 d_S r_D} (\sqrt{T_S}e^{-\Delta/k_BT_S}- \sqrt{T_D}e^{-\Delta/k_BT_D})$ \citep{PeltonenPRB10}, where $r_D$ is the junction resistance per area. The relaxation length in $x$ direction then writes $\ell_{S\_D}=\left(\frac{r_D d_S}{\rho_S} \sqrt{\frac{8}{\pi} \frac{k_BT_D}{\Delta}}\right)^{1/2}=2 \sqrt{r_D}$ $\Omega^{1/2}\mu$m at 300 mK. A small specific resistance $r_D$ is thus needed so that the superconductor is locally efficiently thermalized to the drain. The similar cooling behavior of sample C2 with a direct quasi-particle trap and the original sample C1 leads to the conclusion that heat relaxation in C1 or C2 occurs at a distance from the junction smaller than the distance of the direct trap, \mbox{i.e.} $\ell_{S\_D} \lesssim$ 5 $\mu$m. This statement leads to the estimate $r_D\lesssim10$ $\Omega\mu$m$^2$, which is consistent with expectations based on the fabrication recipe and with a previous evaluation \citep{NguyenNJP}.

Let us now estimate the different terms in the heat balance at the lowest temperature range. We consider sample C1 at its lowest bath temperature of 70 mK, assuming $T_{S}=T_{ph}=T_{bath}$. Joule heating on the normal island, with a resistance of 0.02 $\Omega$ and a typical 1 $\mu$A injection current, is only 20 fW and thus negligible. The electron-phonon coupling power in the Cu island $\dot{Q}_{eph}$ = 0.17 pW is small in comparison to the cooling power $\dot{Q}_{NIS}$ = 20 pW. The phenomenological Dynes factor $\gamma=G_0/G_N$ is the ratio of the conductance at zero bias to its normal state value. It captures the contribution of possible pin holes in the tunnel barrier, inverse proximity effect in the superconductor and the effect of the environment on electron tunneling. The fit of the current-voltage characteristics (data not shown) gives $\gamma=2.5\times 10^{-4}$, which brings a related parasitic heating power $\Delta^2\gamma/R_T\simeq$ 8 pW. As $T_{N,min}\simeq2.5\gamma^{2/3}T_c$ \citep{PekolaPRL04}, the observed minimum temperature $T_{N,min} \simeq$ 30 mK could be related to a $\gamma$ parameter at least three times larger than the measured value. Thus, the Dynes smearing of the superconductor density of states cannot account for the observed temperature saturation.

The discrepancy in the power balance necessarily comes from other sources that we do not have a direct way to probe. Besides potential candidates such as phonon heating \citep{PascalPRB11} or near field heat transport \citep{Biehs}, we also suspect that the non-uniformity of the superconducting gap contributes to the saturation of $T_{N,min}$. It is well-known that the superconducting gap $\Delta$ of a thin Al film can have different values depending on the fabrication details. The gap can be tuned by the grain size of the film \citep{Yamamoto,Aumentado}, and different crystal orientations can have a different value up to $3\%$ \citep{WellsPRB, BiondiPRB}. To model the performance of the cooler, we assume a gaussian distribution of $\Delta$ with a standard deviation $\delta\Delta$ and calculate the maximum cooling power $\dot{Q}_{NIS}^{opt}$ normalized to its value at $\delta\Delta=0$. Figure 4b shows our result at different temperatures. With $\delta\Delta/\Delta=3\%$ at 30 mK ($T/T_c=0.03$), the cooling power would be reduced to one half of its nominal value. 

\section{Conclusions}
In conclusion, accumulation of hot quasi-particles in the superconducting leads limits the cooler performance, even with an efficient quasi-particle drain. We reduced this effect by tuning the tunnel barrier, and demonstrated that such a cooler reaches a 32 mK electron temperature from a 150 mK bath temperature and performs outstandingly over a wide range of temperatures. At temperatures where electron-phonon interaction is strong, phonon cooling in the suspended normal metal island boosts the performance of the cooler. This refrigerator has a significant power and can be easily integrated, so that it opens new possibilities to cool practical devices to the sub 50
mK regime.

\section{Acknowledgments}
We have benefited from discussions with D. V. Averin, Y. A. Pashkin, D. S. Golubev, M. Prunnila, V. F. Maisi, I. Khaymovich, T. T. Heikkil\"a, V. J. Kauppila, and C. B. Winkelmann. We acknowledge the support of the European Community Research Infrastructures under the FP7 Capacities Specific Programme, MICROKELVIN project number 228464, the EPSRC grant EP/F040784/1, the Academy of Finland through its LTQ CoE grant (project no. 250280), and the Cryohall infrastructure. Samples are fabricated in the Micronova Nanofabrication Center of Aalto University.


\begin{thebibliography}{33}
\bibitem{footnoteA2} Because of the large junction size, A2 has 1 nW cooling power at 300 mK. Still, it performs equally well compared to A1 or A3.
\bibitem{footnoteA4} The crossing of data from sample A4 with those from other samples is vaguely inline with expectation as A4 has the smallest metal volume $\mathcal{V}$. Its higher $T_{N,min}$ is presumably due to Joule heating because of the thinner Cu layer.
\bibitem{footnote1} Due to the weak dependence $\ell_D\propto 1/\sqrt{\Sigma_D\rho_D}$, the corresponding numbers for Cu and Au are similar. There is no advantage of replacing AlMn by another material.
\end{thebibliography}
\end{document}